\documentstyle[12pt]{article}

\newcommand{\resection}[1]{\setcounter{equation}{0}\section{#1}}

\def\C {\hbox{\BB C}}

\def\s {\sigma}
\def\be{\begin{equation}}
\def\ee{\end{equation}}
\def\EQ{\begin{equation}}
\def\EN{\end{equation}}
\def\bea{\begin{eqnarray}}
\def\eea{\end{eqnarray}}
\def\to{\rightarrow}
\def\goto{\longrightarrow}
\def\sa{\hspace{0.1in}}

\def\sb{\hspace{0.2in}}
\def\sm{\hspace{0.15in}}

\def\C{{\cal C}}
\def\O{{\cal O}}

\def\D{\Delta}

\begin{document}
\oddsidemargin 5mm
\setcounter{page}{0}
\renewcommand{\thefootnote}{\arabic{footnote}}
\newpage
\setcounter{page}{0}
\begin{titlepage}
\begin{flushright}
ISAS/EP/93/42
\end{flushright}
\vspace{0.5cm}
\begin{center}
{\large {\bf On the Operator Content of the Sinh-Gordon Model }} \\
\vspace{1.5cm}
{\bf
A. Koubek$^1$,
G. Mussardo$^{1,2}$} \\
\vspace{0.8cm}
$^1${\em International School for Advanced Studies, Via Beirut 2-4,
34013 Trieste, Italy} \\
$^2${\em Istituto Nazionale di Fisica Nucleare, Sezione di Trieste}\\
\end{center}
\vspace{6mm}
\begin{abstract}

\noindent
We classify the operator content of local hermitian scalar operators in the
Sinh-Gordon model by means of independent solutions of the form-factor
bootstrap equations. The corresponding linear space is organized into a
tower-like structure of dimension $n$ for the form factors $F_{2n}$ and
$F_{2n-1}$. Analyzing the cluster property of the form factors, a particular
class of these solutions can be identified with the matrix elements of the
operators $e^{k g\phi}$. We also present the complete expression of the form
factors of the elementary field $\phi(x)$ and the trace of the energy-momentum
tensor $\Theta(x)$.
\end{abstract}
\vspace{5mm}
\end{titlepage}
\newpage

\resection{Introduction}

The identification of local operators in a given model and the computation of
their multi-point correlators is one of the most important problems in Quantum
Field Theory (QFT). Although a lot is known about two-dimensional massless QFT
\cite{BPZ,Cardy,ISZ}, our present understanding is incomplete for massive
QFT. A basic principle that allows us to identify the operator content, to
predict the mass spectrum of bound and resonance states and to compute the
correlation functions of the local fields is still missing. Some progress can
be achieved for those massive QFT where the dynamics can be analyzed to a
certain extent by means of non-perturbative methods: this is the case for
instance of two-dimensional integrable massive QFT, mainly characterized by
the factorizable $S$-matrix of the asymptotic states \cite{ZZ,Zam,Musrep}.
In this paper we propose the use of the {\em form factor bootstrap approach}
\cite{Karowski,Smirnov,ZamYL,CM,FMS} to classify the operator content of an
integrable massive QFT. The form factors (FF) $F_n^{\cal O}$ are matrix
elements of a generic local operator ${\cal O}$ between the vacuum and a
set of $n$ particle asymptotic states\footnote{In the following we will
consider a theory with only one particle and we parametrize the momentum
of this particle in terms of the rapidity
$\beta$, $p^0\,=\,m\cosh\beta_i\,\,\, ,
p^1\,=\,m\sinh\beta_i\,\,\,.$},
\EQ
F_n^{\cal O} (\beta_1,\beta_2,\ldots,\beta_n) \,=\,
\langle 0\mid{\cal O}(0,0)\mid \beta_1,\beta_2,\ldots,\beta_n\rangle_{in}
\,\,\, .
\label{FF}
\EN
Based on general properties of a QFT (as unitarity, analyticity and locality),
the form factor bootstrap approach leads to a system of linear functional
equations for the matrix elements $F_n^{\cal O}$. The important point, already
noticed in \cite{CM}, is that these equations do {\em not} refer to any
specific operator. This physical arbitrariness is reflected in the existence
of independent solutions to the same set of functional equations. This
pleasant feature allows us to classify the operator content by associating
each of these solutions to an independent quantum operator of the
theory. In this paper we pursue this idea for the Sinh-Gordon model, which is
particularly suitable for its simple structure as an integrable QFT.

The paper is organized as follows: in Sect.\,2 we discuss general
properties of the Sinh-Gordon model and the form factors of local
operators. In Sect.\,3 we give the solutions of the recursive equations
for the form factors of hermitian scalar operators. The cluster property
of some specific solutions are discussed in Sect.\,4. Our conclusions are
presented in Sect.\,5.

\resection{Form-Factors in the Sinh-Gordon Model}

Let us briefly review the most important properties of the FF in the
Sinh-Gordon theory. For a self-consistent treatment reference is made
to paper \cite{FMS}.

The Sinh-Gordon theory is defined by the action
\EQ
{\cal S}\,=\,\int d^2x \left[
\frac{1}{2}(\partial_{\mu}\phi)^2-\frac{m_0^2}{g^2}
\,\cosh\,g\phi(x)\,\right]\,\,.
\label{Lagrangian}
\EN
It is the simplest example of the affine Toda Field Theories \cite{Toda},
possessing a $Z_2$ symmetry $\phi\rightarrow -\phi$. Its two-particle
$S$-matrix is given by \cite{AFZ}
\EQ
S(\beta,B)\,=\,
\frac{\tanh\frac{1}{2}(\beta-i\frac{\pi B}{2})}
{\tanh\frac{1}{2}(\beta+i\frac{\pi B}{2})} \label{smatrix}\,\, ,
\EN
where $B$ is the so-called renormalized coupling constant
\EQ
B(g)\,=\,\frac{2g^2}{8\pi+g^2} \sm .
\EN
For real values of $g$ the $S$-matrix has no poles in the physical sheet
and hence there are no bound states. The $S$-matrix presents a weak-strong
duality under the transformation $g\rightarrow 8\pi/g$.

The form-factors of the Sinh-Gordon model satisfy the equations relative
to a QFT without bound states \cite{Smirnov} :
\bea
F_n(\beta_1, \dots ,\beta_i, \beta_{i+1},
\dots, \beta_n) &=& F_n(\beta_1,\dots,\beta_{i+1}
,\beta_i ,\dots, \beta_n) \,S(\beta_i-\beta_{i+1}) \,\, ,
\nonumber\\
F_n(\beta_1+2 \pi i, \dots, \beta_{n-1}, \beta_n ) &=& F_n(\beta_2
,\ldots,\beta_{n-1},\beta_n, \beta_1)
\,\, ,\label{axioms}\\
-i\lim_{\tilde\beta \rightarrow \beta}
(\tilde\beta - \beta)
F_{n+2}(\tilde\beta+i\pi,\beta,\ldots,\beta_n) &=&
\left(1-\prod_{i=1}^n S(\beta-\beta_i)\right)\,
F_n(\beta_1,\ldots,\beta_n)\,\,\, .\nonumber
\eea
For an operator ${\cal O}(x)$ of spin $s$, relativistic invariance implies
\EQ
F_n^{\cal O} (\beta_1+\Lambda,\beta_2+\Lambda,\ldots,\beta_n+\Lambda) \,=\,
e^{s\Lambda}\,
F_n^{\cal O} (\beta_1,\beta_2,\ldots,\beta_n) \,\, . \label{relat}
\label{asymp2}
\EN
In the Sinh-Gordon model a convenient parameterization of the $n$-particle
FF, which takes into account their kinematical poles, is given by \cite{FMS}
\EQ
F_n(\beta_1,\ldots,\beta_n)\,=\, H_n\, Q_n(x_1,\ldots,x_n)\,
\prod_{i<j} \frac{F_{\rm min}(\beta_{ij})}{(x_i+x_j)}
\sb ,\label{para}
\EN
where $x_i\equiv e^{\beta_i}$ and $\beta_{ij}=\beta_i-\beta_j$. Let us discuss
the different terms entering the above expression. $F_{\rm min}(\beta)$ is an
analytic function given by
\EQ
F_{\rm min}(\beta,B)\,=\,
\prod_{k=0}^{\infty}
\left|
\frac{\Gamma\left(k+\frac{3}{2}+\frac{i\hat\beta}{2\pi}\right)
\Gamma\left(k+\frac{1}{2}+\frac{B}{4}+\frac{i\hat\beta}{2\pi}\right)
\Gamma\left(k+1-\frac{B}{4}+\frac{i\hat\beta}{2\pi}\right)}
{\Gamma\left(k+\frac{1}{2}+\frac{i\hat\beta}{2\pi}\right)
\Gamma\left(k+\frac{3}{2}-\frac{B}{4}+\frac{i\hat\beta}{2\pi}\right)
\Gamma\left(k+1+\frac{B}{4}+\frac{i\hat\beta}{2\pi}\right)}
\right|^2 \sa .
\EN
It satisfies the functional equations
\EQ
\begin{array}{ccl}
F_{\rm min}(\beta)&=&F_{\rm min}(-\beta)\, S(\beta,B)\,\, ,\\
F_{\rm min}(i\pi-\beta)&=&F_{\rm min}(i\pi+\beta)\,\, , \\
F_{\rm min}(i\pi+\beta,B) F_{\rm min}(\beta,B)&=&
\frac{\sinh\beta}{\sinh\beta+\sinh\frac{i\pi B}{2}}\,\,\, .
\end{array}
\label{Watson2}
\EN
It has a simple zero at the threshold $\beta=0$ (since $S(0,B)=-1$) and no
poles in the physical strip $0\leq {\rm Im}\,\beta \leq \pi$, with an
asymptotic behaviour given by
\EQ
\lim_{\beta \rightarrow \infty} F_{\rm min}(\beta,B) = 1\,\,.
\label{limitefmin}
\EN
$H_n$ are normalization constants, which can be conveniently chosen as
\be
H_{2n+1} = H_{1} \mu^{2n}\sb, \sb H_{2n} = H_{2} \mu^{2n-2}
\,\,\, ,
\ee
with
\be
\mu \equiv
\left ( \frac{4 \sin (\pi B /2)}{F_{min}(i\pi,B)}
\right )^{\frac 12}
\ee
and $H_1$, $H_2$ are two independent parameters. Finally, the functions
$Q_n(x_1,\dots,x_n)$ are symmetric polynomials in the variables $x_i$, which
have to be fixed by the recursion equations satisfied by the form factors.
For FF of spinless operators, their total degree is equal to $n(n-1)/2$. On
the other hand, the partial degree of $Q_n$ in each variable $x_i$ is fixed by
the nature and the asymptotic behaviour of the operator $\cal O$ which is
considered. Exploiting the parameterization (\ref{para}) together with the
functional equations (\ref{axioms}) and (\ref{Watson2}), the polynomials
$Q_n(x_1,\ldots,x_n)$ have to satisfy the recursive equations
\cite{FMS}
\EQ
(-)^n\,Q_{n+2}(-x,x,x_1,\ldots,x_n)\, = \,x  D_n(x,x_1,x_2,\ldots ,x_n)
\,Q_n(x_1,x_2,\ldots,x_n)
\label{rec}
\EN
with
\be D_n(x,x_1,\dots,x_n) = \sum_{k=1}^n \sum_{m=1,odd}^k [m]\, x^{2(n-k)+m}
\sigma_{k}^{(n)}\sigma_{k-m}^{(n)} (-1)^{k+1} \sb .
\label{D_n}
\ee
We have introduced the symbol $[n]$ defined by
\be
[n]\equiv\frac{\sin ( n \frac B2)}{\sin \frac B2}
\ee
and the {\em elementary symmetric polynomials} $\sigma^{(n)}_k(x_1,\dots,x_n)$,
given by the generating function
\cite{Macdon}
\EQ
\prod_{i=1}^n(x+x_i)\,=\,
\sum_{k=0}^n x^{n-k} \,\sigma_k^{(n)}(x_1,x_2,\ldots,x_n)\sb .
\label{generating}
\EN
Conventionally the $\sigma_k^{(n)}$ with $k>n$ and with $n<0$ are zero.
The explicit expressions are given as
\EQ
\s_k^{(n)} = \sum_{i_1 < i_2 <\dots < i_k}^n x_{i_1} x_{i_2} \dots x_{i_k}
\EN
The $\sigma_k^{(n)}$ are linear in each variable $x_i$ and their total degree
is $k$. They satisfy the recursive equations
\EQ
\sigma_k^{(n+2)}(-x,x,x_1,\ldots,x_n)\,=\,
\sigma_k^{(n)}(x_1,x_2,\ldots,x_n)-x^2 \sigma_{k-2}^
{(n)}(x_1,x_2,\ldots,x_n)  \,\,\, .
\label{kinshift}
\EN
As mentioned in the introduction, the linear equations satisfied by the FF do
not refer to any specific operator. Hence the operator content of the model
is encoded in the vector space of solutions of the recursive equations
(\ref{rec}). Among them, we expect to find the FF of the (renormalized)
elementary field $\phi(x)$ and the trace of the energy-momentum tensor
$\Theta$. Their normalizations are given by
\EQ
\begin{array}{lll}
<0 \mid \phi(0)\mid \beta>&=& \frac{1}{\sqrt 2} \,\,\, ,\\
<\beta \mid \Theta(0) \mid \beta> &=& 2\pi m^2 \,\,\,.
\end{array}
\EN
where $m$ is the physical mass.

\resection{Solution of the Recursive Equations}

In this section, we discuss the most general solution of the recursive
equations (\ref{rec}) in the space of symmetric polynomials ${\cal P}$ of
total degree $n(n-1)/2$. Any independent solution defines the matrix elements
of a local scalar operator in the theory. A clarification is in order here. In
the following we will only consider the FF of {\em irreducible operators}.
Their form factors have the distinguished property that they cannot be further
factorized in terms of the elementary symmetric polynomial $\sigma_k^{(n)}$.
We will not consider for instance FF of {\em derivative scalar operators}
of the kind $(\partial \bar \partial)^k \O$. The reason for this is that they
can be simply obtained by multiplying the FF $F_n^{\O}$ by the term
$(\frac {\s_{n-1} \s_{1}}{\s_{n}})^k$.

\subsection{General Solution for $Q_3$ and $Q_4$}

To understand the structure of the linear space of the FF, it is worth
considering the first polynomials $Q_1,..,Q_4$.

By relativistic invariance, $Q_1$ has to be a constant which we
denote by $A_1^{(1)}$. For the same reason $Q_2$ is proportional to $\sigma_1$,
since this is the only symmetric polynomial of degree 1, and we define
$Q_2 = A_1^{(2)} \sigma _1$.

At level 3, the most general symmetric polynomial of total degree 3
is given by
\be
{\cal P}_3=A_1^{(3)} \sigma_3 + A_2^{(3)} \sigma_1 \sigma_2 +A_3^{(3)}
\s _1^3\,\, ,
\ee
and in order to be identified as a form factor of the theory it should satisfy
the recursive equation
\EQ
- {\cal P}_3(-x,x,x_1)= x D_1(x) Q_1(x_1) = - x^2 \sigma_1^{(1)} A_1^1\,\,\, .
\label{upp}
\EN
The solution of (\ref{upp}) gives rise to the most general form factor $Q_3$
\EQ
Q_3\,=\,(A_1^{(1)}-A_2^{(3)}) \sigma_3 + A_2^{(3)} \sigma_1\sigma_2\,\,\, .
\EN
Given $A_1^{(1)}$, the linear space of FF at level 3 is a one-dimensional
manifold parametrized by $A_2^{(3)}$, for instance.

An analogous result holds for $Q_4$. Starting with the most general
symmetric polynomial of total degree 6
\bea
{\cal P}_4&=&A_1^{(4)} \s_4\s_2+A_2^{(4)}\s_3\s_2\s_1+A_3^{(4)} \s_4\s_1^2+
A_4^{(4)}\s_3^2+A_5^{(4)}\s_3\s_1^3 \nonumber\\
&+&A_6^{(4)}\s_2^3+A_7^{(4)}\s_2^2\s_1^2+
A_8^{(4)}\s_2\s_1^4+A_9^{(4)}\s_1^6 \,\,\, ,
\eea
and imposing the recursive equation (\ref{rec}) with initial condition
$A_1^{(2)}$, the final form of $Q_4$ is given by
\EQ
Q_4 = A_2^{(4)}\s_3\s_2\s_1+A_3^{(4)}( \s_4\s_1^2+\s_3^2) \sm ,
\EN
with
\[
A_3^{(4)}+A_2^{(4)}\,=\,A_1^{(2)} \,\,\, .
\]
Also in this case the linear space of FF at level 4 is a one-dimensional
manifold.

It is interesting to note that both solutions can be written as a sum
of determinants\footnote{We use the notation
$\vert\vert A\vert\vert\equiv{\rm det A}$.}
\begin{eqnarray*}
Q_3 &=& A_1^{(3)} \,\left\vert\left\vert
\begin{array}{cc}
0 & \s_3 \\ -1 & 0
\end{array}
\right \vert \right \vert \,
+ \,A_2^{(3)} \,\left\vert\left\vert
\begin{array}{cc}
\s_1 &[2] \s_3\\ 0 & \s_2
\end{array}
\right \vert\right \vert\sb , \\
Q_4 &=& A_1^{(4)} \,\left\vert\left\vert
\begin{array}{ccc}
\s_1 &[2] \s_3 & 0 \\ 0 & \s_2 &[2] \s_4 \\ 0 & 0 & \s_3
\end{array}
\right \vert \right \vert
\,+\, A_2^{(4)} \,\left\vert\left\vert
\begin{array}{ccc}
[2] \s_1 &[3] \s_3 & 0 \\ 1 &[2]\s_2 &[3] \s_4 \\ 0 & \s_1 &[2] \s_3
\end{array}
\right \vert  \right \vert \sb .
\end{eqnarray*}
Let us discuss some properties of $Q_1,\ldots,Q_4$. First of all at any
step of the recursion process a new free parameter enters the solution.
Secondly, the partial degree\footnote{Note that the partial degree in
each $x_i$ determines the asymptotic behaviour of the FF and is fixed by
the total number of $\sigma^{(n)}$'s in each term of the sum.} of $Q_n$
($n=1,\ldots,4$) does not exceed $n-1$. Hence, all these FF will be at the
most constant for $\beta_i \to \infty$. Finally, seeking solutions of the FF
equations which vanish as $\beta_i\to\infty$, there is only a {\em unique}
class of polynomials, i.e. $Q_2$ and $Q_4$ identically zero and $Q_3
\sim \s_3$. In \cite{FMS} it has been shown that these are the first
form-factors of the elementary field $\phi$.

\subsection{Properties of the General Solution $Q_n$}

Important properties of the polynomials $Q_n$ can be easily obtained by
analyzing the recursive equations (\ref{rec}). As a first step let us show
that the partial degree of the polynomials $Q_n$ satisfies
\EQ
{\rm deg}\,(Q_n) \leq n-1\,\,\, .
\EN
We have seen above that this property is true for $Q_1,..,Q_4$. To prove that
this persists for higher polynomials, the cases (a) $Q_n \neq 0$ and (b)
$Q_n =0$ have to be considered separately.

\begin{itemize}
\item
In the case (a) the proof is done by induction. Let us assume
${\rm deg}\,(Q_n)\leq n-1$. Since $D_n$ is bilinear in $\sigma^{(n)}$
(see eq.\,(\ref{D_n})), the partial degree of $Q_{n+2} (-x,x,x_1,\dots x_n)$
in the variables $x_1,\dots x_n$ is smaller or equal to $n+1$. But the partial
degree of $Q_{n+2}(x_1,x_2,\dots x_{n+2})$ is equal to
$Q_{n+2}(-x,x,x_1\dots,x_n)$, therefore the partial degree
of $Q_{n+2}$ must be less or equal to $n+1$.
\item
In the case (b), the space of the solutions is given by the kernel
of eq.\,(\ref{rec}), i.e.
\EQ
Q_{n+2}(-x,x,\dots,x_{n+2}) =0 \,\,\, .
\EN
In the space of polynomials ${\cal P}$ of total degree $\frac{(n+2)(n+1)}{2}$,
there is only one solution of this equation, i.e.
\EQ
Q_{n+2} = \prod_{i < j}^{n+2} (x_i + x_j) \,\,\, .
\label{kernel}
\EN
This polynomial has partial degree $n+1$ and coincides with the denominator
of eq.\,(\ref{para}).
\end{itemize}
We have therefore shown that the partial degree of $Q_n$ must be less or equal
to $(n-1)$ for any spinless irreducible operator. A first consequence
of this statement is that the FF of such operators cannot diverge in the limit
$\beta_i\to\infty$. Secondly, it is now easy to understand the appearance of
one additional parameter at each step of the recursion process. This is simply
because the total dimension of the space of the polynomials $Q_n$ is given by
the dimension of the space of the polynomial $Q_{n-2}$, summed with the
dimension of the kernel. Since the kernel is a one-dimensional manifold, the
dimension of the space of solutions increases exactly by one at each step of
the recursion. With the initial conditions ${\rm dim}\,(Q_1)
={\rm dim}\,(Q_2)=1$, we finally obtain
\be
{\rm dim}\,(Q_{2n-1}) = {\rm dim}\,(Q_{2n})=n \sb .
\ee
Therefore the most general FF at level $n$ of irreducible scalar operators
span a finite linear space which can be expressed in terms of a basis $Q_n^k$
\bea
Q_{2n}(A_1^{(2n)},\ldots,A_n^{(2n)}) &=&
\sum_{p=1}^{n} A_p^{(2n)} Q_{2n}^p \label{ansatz} \\
Q_{2n-1}(A_1^{(2n-1)},\ldots,A_n^{(2n-1)}) &=&
\sum_{p=1}^{n} A_p^{(2n-1)} Q_{2n-1}^p \nonumber \,\,\, .
\eea
Each of these polynomials defines the matrix elements of a quantum operator of
the Sinh-Gordon model. Note that the dimension of the linear space of the
FF grows exactly as the number of powers of the elementary field $\phi(x)$,
i.e. $\phi ^k$, $k < n$. Therefore we expect to identify the tower structure
of the FF with the space of the matrix elements of the composite operators
$\phi ^k$. The precise correspondence will be discussed in a forthcoming
publication \cite{future}.

\subsection{Elementary Solutions}

An interesting class of solutions of the recursive equations (\ref{rec})
from which we can extract a basis for the FF space is given by\footnote{For
simplicity we suppress the dependance of $Q_n(k)$ on the variables
$x_i$.}
\be
Q_{n}(k) = \vert\vert M_{ij}(k) \vert\vert \sm ,
\ee
where $M_{ij}(k)$ is an $(n-1)\times (n-1)$ matrix with entries
\be M_{ij}(k) =
\s_{2i-j}\, [i-j+k] \sm .
\label{element}
\ee
These polynomials, which we call {\em elementary solutions}, depend on
an arbitrary integer $k$ and satisfy
\be
Q_{n}(k)\,=\, (-1)^{n+1} Q_{n}(-k) \label{pr1} \sm .
\ee
Although all $Q_n(k)$ are solutions of (\ref{rec}) not all of them
can be linearly independent. The reason is that the dimension of space of
the solutions at level $N=2n$ (or $N=2n-1$) is $n$ at most. The first non
trivial $Q_n(k)$ are given by the determinant
\be
Q_3(k)\, =\,
\left \vert\left\vert \begin{array}{cc}
[k] \s_1 & [k+1] \s_3  \\ \mbox{[k-1]} &   [k] \s_2 \end{array} \right \vert
\right\vert \sm .
\ee
Using the trigonometrical identity $[n]^2-[n-1][n+1]=1$, it is easy to see
that they satisfy eq.\,(\ref{rec}) (with $A_0^1=1$) for any integer
value of $k$.

The whole set\footnote{The first representatives of them were computed in
\cite{FMS}.} of FF of the elementary field $\phi(x)$ and the trace of the
energy-momentum tensor $\Theta(x)$ can be easily expressed in terms
of the $Q_n(k)$. In fact the FF of the field $\phi(x)$ are
given by $Q_n(0)$. Note that the FF of $\phi(x)$ are automatically zero for
even $n$, in agreement with the $Z_2$ parity of the model. On the contrary,
for odd $n$ they vanish asymptotically when $\beta_i\rightarrow
\infty$, as follows from the LSZ reduction formula \cite{FMS}. Concerning
the FF of $\Theta(x)$, these are given by the even polynomials $Q_{2n}(1)$,
which tend to a constant when $\beta_i\rightarrow \infty$.

\resection{Cluster-Property and Exponential Operators}

We can associate a quantum operator $\Psi_k(x)$ to any elementary solution
$Q_n(k)$ ($k\neq 0$). To identify such operators it is interesting to analyze
the cluster property of their FF. By cluster transformation we generally mean
the behaviour of a form factor under the shift of a subset of the rapidities,
i.e.
\EQ
F_n^{\O_a}(\beta_1+\Delta,\dots,\beta_m+\Delta,\beta_{m+1},\dots,
\beta_{n}) \,\,\, .
\label{cluss}
\EN
Taking the limit $\Lambda\rightarrow\infty$, $F_n^{{\cal O}_a}$ can be
decomposed into two functions of $m$ and $(n-m)$ variables respectively.
It is easy to prove that both functions satisfy all the set of axioms
(\ref{axioms}). Therefore they can be considered as FF of some operators
${\cal O}_b$ and ${\cal O}_c$
\be
\lim_{\Lambda\rightarrow \infty}
F_n^{\O_a}(\beta_1+\Delta,\dots,\beta_m+\Delta,\beta_{m+1},\dots,
\beta_{n})
= F_m^{\O_b}(\beta_1,\dots,\beta_m)
F_{n-m}^{\O_c}(\beta_{m+1},\dots,\beta_n)
\ee
Shortly,
\[
{\cal O}_a\, \rightarrow \,{\cal O}_b \times {\cal O}_c \,\,\, .
\]
We will prove that the operators $\Psi_k$ are mapped onto themselves
under the cluster transformation, i.e.
\EQ
\Psi_k \,\rightarrow \, \Psi_k \times \Psi_k \,\,\, .
\EN
To this aim, let us introduce some notations.

It is useful to define the cluster-operator $\C_m$ (acting on the symmetric
functions) by means of
\be
\C_m \left ( f(x_1,\dots,x_n) \right ) \equiv f(x_1e^\D,x_2e^\D,\dots,
x_me^\D,x_{m+1},\dots,x_n ) \sa m<n \sa .
\ee
For example,
\be \C_1(\s_1^{(n)})=e^\D x_1+x_2\dots+x_n = e^\D \s_1^{1}+ \hat \s_1^{(n-1)}
\sm ,\ee
where we have defined
$$
\hat \s_i^{(n-k)} \equiv \s_i^{(n-k)} (x_{n-k+1},x_{n-k+2},\dots,x_n) \sb.
$$
One easily finds that
\be
\C_m(\s_k^{(n)})= \sum_{i=1}^{k} \s_{k-i}^{(m)} e^{(k-i)\D} \hat \s_i^{(n-m)}
\sb .\label{csigma}
\ee
Since the cluster properties are fixed by the leading term of this sum,
we have
\be
\begin{array}{ll}
\C_m(\s_k^{(n)})\sim \s_m^{(m)} e^{m\D} \hat \s_{k-m}^{(n-m)} & m \leq
k \sm ,\\
\C_m(\s_k^{(n)})\sim \s_k^{(m)} e^{k\D} & m \geq k \sm .
\end{array}
\label{clussym}
\ee
Now let us consider separately the cluster property of each
term entering their parametrization
\be
F_n^k(\beta_1,\dots,\beta_n) = H_n^k \,Q_n(k)\,
\prod_{i<j}^n \frac{F_{\rm min}(\beta_{ij})}{(x_i+x_j)}  \label{fn}\,\,\, .
\ee
{}From eq.\,(\ref{limitefmin}), we have
\be
\prod _{i<j}^n F_{\rm min}(\beta_{ij}) \goto
\prod _{i<j}^m F_{\rm min}(\beta_{ij})
\prod _{i<j=m+1}^n F_{\rm min}(\beta_{ij})\,\,\,.
\label{first}
\ee
Using eq.\,(\ref{clussym}), the cluster property of the elementary solution
$Q_n(k)$ is given by
\be
\C_m(Q_{n}(k) ) \sim  h(n,m)\,[k]\,Q_{m}(k)\, Q_{n-m}(k)\sm ,
\label{cq}
\ee
where
$$
h(n,m)=e^{\D m (n-\frac{m+1}2)}\,(\s_m^{(m)})^{n-m} \,\,\,.
$$
Concerning the denominator in (\ref{fn}), we write it initially as
\be
\prod_{i<j}^{n} (x_i+x_j) = \vert\vert \s_{2i-j}^{(n)}
\vert\vert^{\langle n-1\rangle}\sb ,
\ee
where the index $\langle n-1 \rangle$ indicates the dimension of the matrix.
Using eq.\,(\ref{clussym}), we obtain
\be
\C_m\left ( \vert\vert \s_{2i-j}^{(n)} \vert \vert^{\langle n-1\rangle}
\right ) \sim
h(n,m)\, \vert \vert \s_{2i-j}^{(m)}
\vert \vert^{\langle m-1\rangle}
\, \times \, \vert \vert\hat \s_{2i-j}^{(n-m)}\vert
\vert^{\langle n-m-1\rangle}\,\,\, .
\label{cden}
\ee
Choosing the normalization constants $H_1^k$ and $H_2^k$ as
\be
H_1^k =\mu [k] \sa,\sa H_2^k=\mu^2\, [k] \sm  ,\label{h12}
\ee
and using eqs.\,(\ref{first}), (\ref{cq}) and (\ref{cden}), we conclude that
the FF of $\Psi_k$ are mapped onto themselves under the cluster transformation.
Since this is a distinguished property of exponential operators \cite{Smirnov},
it is natural to identify the operators $\Psi_k$ with the fields
$e^{k g \phi}$. A non-trivial check is given by the computation of their
anomalous dimensions $\Delta_k(g)$ obtained in the ultraviolet limit. The
quantity $\Delta_k(g)$ can be computed by analyzing the limit
$x\rightarrow 0$ of the correlation functions
\begin{eqnarray}
& &G_{k,m}(x)\,=\,\langle\Psi_k(x)\,\Psi_m(0)\rangle\,=
\\
& &\hspace{3mm} =\,\sum_{n=0}^{\infty}
\int \frac{d\beta_1\ldots d\beta_n}{n! (2\pi)^n}
F_n^{\Psi_k}(\beta_1\ldots \beta_n)
F_n^{\Psi_m}(\beta_n\ldots \beta_1)
\exp \left(-mr\sum_{i=1}^n\cosh\beta_i
\right) \,\,\, .\nonumber
\end{eqnarray}
At first order in $g$, we find $\Delta_k(g)=-k^2 g^2/8\pi$ which
coincides with the anomalous dimensions of the exponential operators
$e^{k\,g\,\phi(x)}$, computed from conformal perturbation theory.

\resection{Conclusions}

\,

The computation of Green functions and the classification of the operator
content are key problems in Quantum Field Theory. For integrable models, a
promising development has been achieved by the bootstrap approach.

In this paper we have given the general form factors of
scalar irreducible operators in the Sinh-Gordon model.
We have found an infinite number of elementary solutions $Q_n(k)$. As far
as the renormalized coupling constant $B(g)$
is irrational, they define the FF of the independent operators $\Psi_k$.
This is no longer true if we take rational values of $B(g)$. In these cases
an identification of different operators occurs due to the symmetry
of the quantum-symbol $[n]$ and the space of the fields $\Psi_k$ is
drastically reduced. This reflects the situation occurring in conformal field
theory and confirms the hope that one may find similar finite-dimensional
representations for the symmetry algebra of massive integrable QFT at
rational values of the coupling constant.

\vspace{1cm}

\section*{Acknowledgments}

We are grateful to J.L. Cardy, G. Delfino, A. Fring, A. Schwimmer,
P. Simonetti and F. Smirnov for useful discussions.

\end{document}